\documentclass[aps,prl,twocolumn,showpacs,superscriptaddress,amsmath,amssymb]{revtex4}

\usepackage{graphicx}% Include figure files
\usepackage{dcolumn}% Align table columns on decimal point
\usepackage{bm}% bold math

\def\D0{D\O} 

\begin{document}

\preprint{Version 3.2}

\title{Search for charged Higgs bosons in decays of top quarks \\in $p\bar{p}$ collisions at $\sqrt{s}$ = 1.96 TeV}

\affiliation{Institute of Physics, Academia Sinica, Taipei, Taiwan 11529, Republic of China} 
\affiliation{Argonne National Laboratory, Argonne, Illinois 60439} 
\affiliation{University of Athens, 157 71 Athens, Greece} 
\affiliation{Institut de Fisica d'Altes Energies, Universitat Autonoma de Barcelona, E-08193, Bellaterra (Barcelona), Spain} 
\affiliation{Baylor University, Waco, Texas  76798} 
\affiliation{Istituto Nazionale di Fisica Nucleare Bologna, $^v$University of Bologna, I-40127 Bologna, Italy} 
\affiliation{Brandeis University, Waltham, Massachusetts 02254} 
\affiliation{University of California, Davis, Davis, California  95616} 
\affiliation{University of California, Los Angeles, Los Angeles, California  90024} 
\affiliation{University of California, San Diego, La Jolla, California  92093} 
\affiliation{University of California, Santa Barbara, Santa Barbara, California 93106} 
\affiliation{Instituto de Fisica de Cantabria, CSIC-University of Cantabria, 39005 Santander, Spain} 
\affiliation{Carnegie Mellon University, Pittsburgh, PA  15213} 
\affiliation{Enrico Fermi Institute, University of Chicago, Chicago, Illinois 60637}
\affiliation{Comenius University, 842 48 Bratislava, Slovakia; Institute of Experimental Physics, 040 01 Kosice, Slovakia} 
\affiliation{Joint Institute for Nuclear Research, RU-141980 Dubna, Russia} 
\affiliation{Duke University, Durham, North Carolina  27708} 
\affiliation{Fermi National Accelerator Laboratory, Batavia, Illinois 60510} 
\affiliation{University of Florida, Gainesville, Florida  32611} 
\affiliation{Laboratori Nazionali di Frascati, Istituto Nazionale di Fisica Nucleare, I-00044 Frascati, Italy} 
\affiliation{University of Geneva, CH-1211 Geneva 4, Switzerland} 
\affiliation{Glasgow University, Glasgow G12 8QQ, United Kingdom} 
\affiliation{Harvard University, Cambridge, Massachusetts 02138} 
\affiliation{Division of High Energy Physics, Department of Physics, University of Helsinki and Helsinki Institute of Physics, FIN-00014, Helsinki, Finland} 
\affiliation{University of Illinois, Urbana, Illinois 61801} 
\affiliation{The Johns Hopkins University, Baltimore, Maryland 21218} 
\affiliation{Institut f\"{u}r Experimentelle Kernphysik, Universit\"{a}t Karlsruhe, 76128 Karlsruhe, Germany} 
\affiliation{Center for High Energy Physics: Kyungpook National University, Daegu 702-701, Korea; Seoul National University, Seoul 151-742, Korea; Sungkyunkwan University, Suwon 440-746, Korea; Korea Institute of Science and Technology Information, Daejeon, 305-806, Korea; Chonnam National University, Gwangju, 500-757, Korea} 
\affiliation{Ernest Orlando Lawrence Berkeley National Laboratory, Berkeley, California 94720} 
\affiliation{University of Liverpool, Liverpool L69 7ZE, United Kingdom} 
\affiliation{University College London, London WC1E 6BT, United Kingdom} 
\affiliation{Centro de Investigaciones Energeticas Medioambientales y Tecnologicas, E-28040 Madrid, Spain} 
\affiliation{Massachusetts Institute of Technology, Cambridge, Massachusetts  02139} 
\affiliation{Institute of Particle Physics: McGill University, Montr\'{e}al, Qu\'{e}bec, Canada H3A~2T8; Simon Fraser University, Burnaby, British Columbia, Canada V5A~1S6; University of Toronto, Toronto, Ontario, Canada M5S~1A7; and TRIUMF, Vancouver, British Columbia, Canada V6T~2A3} 
\affiliation{University of Michigan, Ann Arbor, Michigan 48109} 
\affiliation{Michigan State University, East Lansing, Michigan  48824}
\affiliation{Institution for Theoretical and Experimental Physics, ITEP, Moscow 117259, Russia} 
\affiliation{University of New Mexico, Albuquerque, New Mexico 87131} 
\affiliation{Northwestern University, Evanston, Illinois  60208} 
\affiliation{The Ohio State University, Columbus, Ohio  43210} 
\affiliation{Okayama University, Okayama 700-8530, Japan} 
\affiliation{Osaka City University, Osaka 588, Japan} 
\affiliation{University of Oxford, Oxford OX1 3RH, United Kingdom} 
\affiliation{Istituto Nazionale di Fisica Nucleare, Sezione di Padova-Trento, $^w$University of Padova, I-35131 Padova, Italy} 
\affiliation{LPNHE, Universite Pierre et Marie Curie/IN2P3-CNRS, UMR7585, Paris, F-75252 France} 
\affiliation{University of Pennsylvania, Philadelphia, Pennsylvania 19104}
\affiliation{Istituto Nazionale di Fisica Nucleare Pisa, $^x$University of Pisa, $^y$University of Siena and $^z$Scuola Normale Superiore, I-56127 Pisa, Italy} 
\affiliation{University of Pittsburgh, Pittsburgh, Pennsylvania 15260} 
\affiliation{Purdue University, West Lafayette, Indiana 47907} 
\affiliation{University of Rochester, Rochester, New York 14627} 
\affiliation{The Rockefeller University, New York, New York 10021} 
\affiliation{Istituto Nazionale di Fisica Nucleare, Sezione di Roma 1, $^{aa}$Sapienza Universit\`{a} di Roma, I-00185 Roma, Italy} 

\affiliation{Rutgers University, Piscataway, New Jersey 08855} 
\affiliation{Texas A\&M University, College Station, Texas 77843} 
\affiliation{Istituto Nazionale di Fisica Nucleare Trieste/Udine, I-34100 Trieste, $^{bb}$University of Trieste/Udine, I-33100 Udine, Italy} 
\affiliation{University of Tsukuba, Tsukuba, Ibaraki 305, Japan} 
\affiliation{Tufts University, Medford, Massachusetts 02155} 
\affiliation{Waseda University, Tokyo 169, Japan} 
\affiliation{Wayne State University, Detroit, Michigan  48201} 
\affiliation{University of Wisconsin, Madison, Wisconsin 53706} 
\affiliation{Yale University, New Haven, Connecticut 06520} 
\author{T.~Aaltonen}
\affiliation{Division of High Energy Physics, Department of Physics, University of Helsinki and Helsinki Institute of Physics, FIN-00014, Helsinki, Finland}
\author{J.~Adelman}
\affiliation{Enrico Fermi Institute, University of Chicago, Chicago, Illinois 60637}
\author{T.~Akimoto}
\affiliation{University of Tsukuba, Tsukuba, Ibaraki 305, Japan}
\author{B.~\'{A}lvarez~Gonz\'{a}lez$^q$}
\affiliation{Instituto de Fisica de Cantabria, CSIC-University of Cantabria, 39005 Santander, Spain}
\author{S.~Amerio$^w$}
\affiliation{Istituto Nazionale di Fisica Nucleare, Sezione di Padova-Trento, $^w$University of Padova, I-35131 Padova, Italy} 

\author{D.~Amidei}
\affiliation{University of Michigan, Ann Arbor, Michigan 48109}
\author{A.~Anastassov}
\affiliation{Northwestern University, Evanston, Illinois  60208}
\author{A.~Annovi}
\affiliation{Laboratori Nazionali di Frascati, Istituto Nazionale di Fisica Nucleare, I-00044 Frascati, Italy}
\author{J.~Antos}
\affiliation{Comenius University, 842 48 Bratislava, Slovakia; Institute of Experimental Physics, 040 01 Kosice, Slovakia}
\author{G.~Apollinari}
\affiliation{Fermi National Accelerator Laboratory, Batavia, Illinois 60510}
% ...

\collaboration{CDF Collaboration\footnote{With visitors from $^a$University of Massachusetts Amherst, Amherst, Massachusetts 01003,
$^b$Universiteit Antwerpen, B-2610 Antwerp, Belgium, 
$^c$University of Bristol, Bristol BS8 1TL, United Kingdom,
$^d$Chinese Academy of Sciences, Beijing 100864, China, 
$^e$Istituto Nazionale di Fisica Nucleare, Sezione di Cagliari, 09042 Monserrato (Cagliari), Italy,
$^f$University of California Irvine, Irvine, CA  92697, 
$^g$University of California Santa Cruz, Santa Cruz, CA  95064, 
$^h$Cornell University, Ithaca, NY  14853, 
$^i$University of Cyprus, Nicosia CY-1678, Cyprus, 
$^j$University College Dublin, Dublin 4, Ireland,
$^k$University of Edinburgh, Edinburgh EH9 3JZ, United Kingdom, 
$^l$Universidad Iberoamericana, Mexico D.F., Mexico,
$^m$Queen Mary, University of London, London, E1 4NS, England,
$^n$University of Manchester, Manchester M13 9PL, England, 
$^o$Nagasaki Institute of Applied Science, Nagasaki, Japan, 
$^p$University of Notre Dame, Notre Dame, IN 46556,
$^q$University de Oviedo, E-33007 Oviedo, Spain, 
$^r$Texas Tech University, Lubbock, TX  79409, 
$^s$IFIC(CSIC-Universitat de Valencia), 46071 Valencia, Spain,
$^t$University of Virginia, Charlottesville, VA  22904,
$^u$Bergische Universit\"at Wuppertal, 42097 Wuppertal, Germany,
$^{cc}$On leave from J.~Stefan Institute, Ljubljana, Slovenia, 
}}
\noaffiliation

\date{\today}

\begin{abstract}
We report on the first direct search for charged Higgs 
bosons decaying into $c\bar{s}$ in $t\bar{t}$ events 
produced by $p\bar{p}$ collisions at $\sqrt{s}$ = 1.96 TeV. 
The search uses a data sample corresponding to an integrated 
luminosity of 2.2 ${\rm fb^{-1}}$ collected by the CDF II 
detector at Fermilab and looks for a resonance in the invariant 
mass distribution of two jets in the lepton+jets 
sample of $t\bar{t}$ candidates. We observe no evidence 
of charged Higgs bosons in top quark decays. Hence,
95\% upper limits on the top quark decay 
branching ratio are placed at 
${\cal B}(t \rightarrow H^+ b) <$ 0.1 to 0.3 for
charged Higgs boson masses of 60 to 150 GeV/$c^2$
assuming ${\cal B}(H^+ \rightarrow c\bar{s})$ = 1.0. 
The upper limits on ${\cal B}(t \rightarrow H^+ b)$ 
are also used as model-independent limits on the 
decay branching ratio of top quarks to generic scalar 
charged bosons beyond the standard model.
\end{abstract}


\maketitle
The standard model (SM) is remarkably successful in 
describing the fundamental particles and their 
interactions. Nevertheless, it is an incomplete theory.
An important unresolved question is the 
mechanism of electroweak symmetry breaking (EWSB). In the 
SM, a single complex scalar doublet field breaks the 
symmetry, resulting in massive electroweak gauge bosons and
a single observable Higgs boson~\cite{higgs}. To date, 
the Higgs boson has not been discovered, and 
consequently the mechanism of EWSB remains in question.

Beyond the SM, many diverse hypotheses with extended Higgs
sectors have been proposed to explain EWSB. The simplest
extension is a two Higgs-doublet model (2HDM). The minimal 
supersymmetric standard model (MSSM) employs the 
type-II 2HDM, where at leading order one doublet couples to 
the up-type fermions and the other couples to the 
down-type fermions~\cite{mssm}.  
The two Higgs doublet fields manifest themselves as 
two charged Higgs bosons ($H^\pm$) and three 
neutral Higgs bosons ($h$, $H$, $A$). 

In 2HDM and MSSM, the top quark is allowed 
to decay into a charged Higgs boson ($H^+$)~\cite{ch} 
and a bottom quark. The tree level branching ratio of top 
quarks to $H^+$, ${\cal B}(t \rightarrow H^+ b)$, is 
a function of the $H^+$ mass ($m_{H^+}$) 
and  $\tan\beta$. The parameter $\tan\beta$ is  
the ratio of vacuum expectation values of the 
two Higgs doublets. In MSSM,  
${\cal B}(t \rightarrow H^+ b)$ also depends on 
extra parameters related to the masses and couplings of the 
other supersymmetric particles. The 
${\cal B}(t \rightarrow H^+ b)$ is relatively large if 
$\tan\beta$ is small ($\lesssim$ 1) or large
($\gtrsim$ 15)~\cite{phen}.
At low $\tan\beta$, $H^+$ predominantly decays into 
$c\bar{s}$ for low $m_{H^+}$ ($\lesssim$ 130 
GeV/$c^2$) and $t^* \bar{b}$ ($\rightarrow W b\bar{b}$)
\cite{tstar} for higher $m_{H^+}$. In the high 
$\tan\beta$ region, the $H^+$ decays into 
$\tau^+ \nu$ almost 100\% of the time. 

At Tevatron collider experiments, $H^+$ searches 
have been performed for the $H^+ \rightarrow \tau\bar{\nu}$ 
in $t\bar{t}$ decays.
Some searches placed direct upper limits on
${\cal B}(t \rightarrow H^+ b)$ by taking advantage 
of the expectation that %exclusive branching ratio 
${\cal B}(H^+ \rightarrow \tau\bar{\nu})$ = 1.0 at 
high $\tan\beta$~\cite{paper1}. 
%
%Taking advantage of the exclusive decay 
%branching ratio of $H^+ \rightarrow \tau^+ \nu$
%in the high $\tan\beta$ region, some searches
%set direct limits on ${\cal B}(t \rightarrow H^+ b)$
%
%
Other searches set limits on the MSSM 
parameter plane ($m_{H^+}$, $\tan\beta$)
using inclusive $H^+$ decay branching 
ratios in the MSSM~\cite{paper2}. 
The various $H^+$ final states %result in different 
supplement the SM $t\bar{t}$ decay channels.
The previous searches focused on measuring deviations 
from the SM prediction for the $t\bar{t}$ production 
and decay, rather than reconstructing $H^+$ bosons.

In this Letter, we report on the first direct search for 
$H^+ \rightarrow c\bar{s}$ produced in top quark decays by
fully reconstructing the $c\bar{s}$ mass. The final
state of $H^+ \rightarrow c\bar{s}$ is mostly two 
jets, as is the hadronic decay of the $W$ boson
\cite{w} in SM top quark decays. The search is 
performed by looking for a 
second peak in the dijet mass spectrum (in addition to that
from the $W$ boson) in top quark decays.

In the SM, each top quark decays into a $W$ boson 
and a $b$-quark exclusively. In this analysis we use 
%$t\bar{t}$ candidates in 
the lepton+jets $t\bar{t}$ sample~\cite{lepton}, 
where in the SM one $W$ decays 
to quarks ($q\bar{q}^\prime$) and 
the other $W$ decays to $e \bar{\nu}$ or $\mu \bar{\nu}$.
Each final-state quark is assumed to form a hadronic jet; 
the jets are clustered using a cone algorithm with a cone radius 
$\Delta R$ (= $\sqrt{ (\Delta \eta)^2 + (\Delta \phi)^2 }$)
of 0.4~\cite{jes}. This lepton+jets sample
has a good signal-to-background ratio for $t\bar{t}$ and is 
ideal for dijet mass analysis.

The CDF II experiment at the Fermilab Tevatron 
measures the products of proton-antiproton collisions 
at $\sqrt{s}$ = 1.96 TeV. 
The lepton momentum is measured using an eight-layer 
silicon microstrip detector and a cylindrical drift 
chamber immersed in a 1.4 T magnetic field.
The energies of electrons and jets are measured 
using calorimeters with acceptance up to 
pseudorapidity as of $|\eta|$ = 3.6~\cite{eta}. 
Charged particle detectors outside the calorimeter 
identify muon candidates up to  $|\eta|$ = 1.0. 
Details of CDF II can be found 
elsewhere~\cite{tmt}.

Lepton+jets $t\bar{t}$ events are selected by 
requiring an electron or a muon with $p_T>$ 20 GeV 
within $|\eta|$=1 and by requiring missing transverse 
energy larger than 20 GeV to account for the 
neutrino~\cite{met}.  
Then, the four most energetic jets (called leading jets) 
within $|\eta|<$ 2.0 are required to have $E_T>$ 20 GeV
after jet energy corrections~\cite{jes}. 
In addition, at least two of the leading jets are 
required to contain a long-lived hadron containing a 
$b$-quark~\cite{btag} by demanding that these jets contain 
tracks forming a displaced secondary vertex 
(called a $b$-tag). 

The SM processes are regarded as backgrounds for the
$H^+$ search. The largest background is 
$W$ bosons in SM $t\bar{t}$ events 
(92\% of the total background). The rest of the SM 
processes are referred to as non-$t\bar{t}$ 
backgrounds. These include $W$+jets, multijets, 
$Z$+jets, diboson ($WW$, $WZ$, $ZZ$), and 
single top events. Details of the non-$t\bar{t}$ 
background estimation method are given in~\cite{btag}. 
Assuming a $t\bar{t}$ cross section of 6.7 pb~\cite{x-sec}
and a top quark mass of 175 GeV/$c^2$,
we expect 152.6 $\pm$ 25.0 events from SM $t\bar{t}$ 
production and 13.9 $\pm$ 7.5 events from non-$t\bar{t}$ 
backgrounds in the 2.2 ${\rm fb^{-1}}$ CDF II data sample. 

The mass of the $H^+$ candidate is directly reconstructed 
using the two jets. The mass resolution is improved by 
reconstructing the $t\bar{t}$ event as a whole 
%For the $t\bar{t}$ event reconstruction, we employ 
with a kinematic fitter used for the precision top 
quark mass measurement described in Ref.~\cite{tmt}. 
The original kinematic fitter is modified for the $H^+$ search.
In the fitter, the lepton, the missing $E_T$ 
(from a neutrino), and the four leading jets 
are assigned to the decay particles from the $t\bar{t}$ event,
and the quality of the assignment is evaluated using 
this $\chi^2$:
\begin{eqnarray}
&\chi^{2}& = \sum_{k=jjb,l\nu b} \frac{(M_k-M_t)^{2}}{{\Gamma_{t}}^{2}} 
+ \sum_{i=l,4 {\rm jets}} \frac{(p_T^{i,{\rm fit}}-p_T^{i,{\rm meas}})^{2}}{{\sigma_{i}}^{2}} \nonumber\\
&+& \frac{(M_{l\nu}-M_{W})^{2}}{{\Gamma_{W}}^{2}} 
+ \sum_{j=x,y} \frac{({p_{j}}^{UE,{\rm fit}}-{p_{j}}^{UE,{\rm meas}})^{2} }{{\sigma_{UE}}^{2}}.
\label{eq:chi2}
\end{eqnarray}
The $\chi^2$ is minimized by constraining leptonic 
$W$ final state ($l\nu$) to have the $W$ invariant mass 
(80.4 GeV/$c^2$)~\cite{pdg} and both top quark final states 
($b l \nu$ and $bjj$) to have the same top quark mass 
of 175 GeV/$c^2$. No constraint is imposed on the dijet 
mass of the hadronic boson ($jj$). In the mass constraints, 
the transverse energies of the final-state objects 
($p_T^{i,{\rm meas}}$) are allowed to vary 
within measurement uncertainties ($\sigma_i$).
The unclustered energy ($p_j^{UE,{\rm meas}}$) is the sum of 
measured transverse energies not included in the 
leading jets $E_T$ and is used to correct the missing $E_T$.
In the jet assignment, $b$-tagged jets are assigned to 
the $b$-quarks. The jets assigned to the $b$-quarks are
called $b$-jets, and the other two jets are called h-jets.
If the $t\bar{t}$ event has more than two 
$b$-tagged jets, the jets with the best $\chi^2$ 
are assigned to $b$-quarks. Then, we reconstruct the mass of hadronic 
boson using two h-jets with fit energies ($p_T^{i,{\rm fit}}$). 
In this kinematic event reconstruction, only 55\% 
of the SM $t\bar{t}$ events have correctly 
matching jets. The wrong jet-parton assignments dominantly 
come from hard radiation jets which are selected 
as leading jets and from the falsely $b$-tagged 
jets originating from the hadronic decays of $W$ bosons.

The expected dijet mass distributions of $H^+$ and $W$ 
in top quark decays are produced using the \textsc{pythia} 
generator~\cite{pythia}  and the full CDF II detector simulation.
The \textsc{alpgen} generator~\cite{alpgen} with the \textsc{pythia}
parton shower simulation is used for non-$t\bar{t}$ backgrounds. 
In the simulation sample, the $H^+$ is forced to decay 
solely into $c\bar{s}$ with zero width and with masses 
ranging from 60 to 150 GeV/$c^2$. 

The simulation shows that
the reconstructed $H^+$ has a significant low-mass tail,
which is predominantly caused by final-state gluon 
radiation (FSR) from the hadronic decays of the Higgs
boson. The hard FSR results in 
more than four final-state jets in a lepton+jets 
$t\bar{t}$ event.
To recover the energy loss due to the FSR,
the fifth most energetic jet is merged with the 
closest jet among the four leading ones if the 
pair has a $\Delta R$
distance smaller than 1.0, provided that 
the fifth most energetic jet has 
$E_T>$ 12 GeV and $|\eta|<$2.4.
Merging the fifth jet results in better jet 
energy resolution and improves the $m_{H^+}$ 
resolution by approximately 5\% in 
more than four final jets events 
for 120 GeV/$c^2$ Higgs sample.

\begin{figure}[htbp]
\includegraphics[width=3.375in]{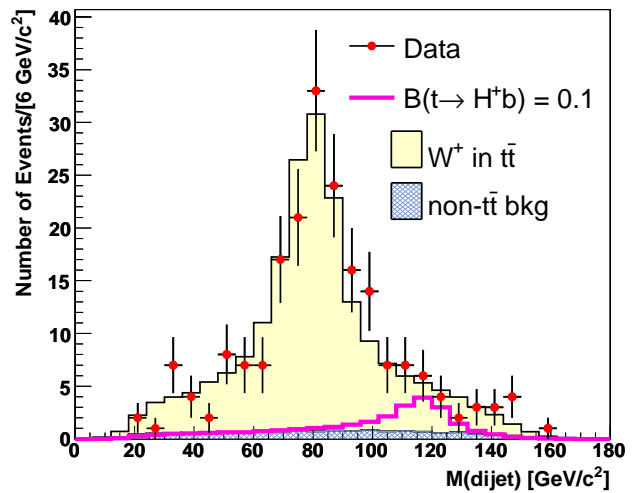}
\caption{\label{fig:mass}  Observed dijet mass 
distribution (crosses) compared with  
background distributions of 
$W$ bosons (filled) and non-$t\bar{t}$ processes 
(cross hatched) in CDF II data sample of 2.2 
${\rm fb^{-1}}$; the background distributions 
are added on top of each other.
An example of the dijet mass distribution 
from 120 GeV/${\rm c^2}$ $H^+$ bosons (bold line) 
is overlaid assuming ${\cal B}(t \rightarrow H^+ b)$
= 0.1, which is about the 95\% C.L. upper limit on 
${\cal B}(t \rightarrow H^+ b)$.}
\end{figure}

In the CDF II data sample of 2.2 ${\rm fb^{-1}}$, 
we observe 200 $t\bar{t}$ candidates in the lepton+jets
decay channel. No significant excess is observed
in the dijet invariant mass of top quark decays. 
Figure~\ref{fig:mass} shows that the observed dijet mass 
distribution agrees with the SM expectations.
Hence, we extract upper limits on 
${\cal B}(t \rightarrow H^+ b)$ using a binned 
likelihood fit on the dijet mass distribution. 

The binned likelihood (LH) function is constructed 
employing Poisson probabilities:
\begin{equation}
{\cal L} = \prod_i \frac{\nu_{i}^{n_{i}} \times e^{-\nu_{i}} }{n_{i}!} \times G({N_{\rm bkg}},\sigma_{N_{\rm bkg}}).
%\label{eq:lh}
\end{equation}
The probability of finding events in the mass
bin $i$ comes from a set of simulated dijet mass 
distributions of  $H^+$, $W$, and non-$t\bar{t}$ 
backgrounds. These distributions are called templates. The 
Poisson probability ($P^i$) in each bin is computed 
from the number of observed events, $n_i$, and from 
the number of expected events, 
$\nu_i = P_{H^+}^i \times N_{H^+} + P_{W}^i \times N_{W} + P_{\rm bkg}^i \times N_{\rm bkg}$, where $N_{H^+}$, 
$N_{W}$, and $N_{\rm bkg}$ are parameters representing 
the total number of events in each template category.
The minimization of $- \ln {\cal L}$ gives the most 
probable values for $N_{H^+}$, $N_{W}$, and $N_{\rm bkg}$.
In the LH fit, $N_{H^+}$ and $N_{W}$ are free to vary,
however, the non-$t\bar{t}$ background 
($N_{\rm bkg}$) is estimated independently  
and is allowed to vary within 
its Gaussian uncertainty ($\sigma_{\rm N_{bkg}}$).  
Based on the number of events from the LH fit, a 
${\cal B}(t \rightarrow H^+ b)$ is extracted assuming
${\cal B}(H^+ \rightarrow c\bar{s})$ = 1.
In Figure~\ref{fig:mass}, dijet mass 
distributions of the SM events
are normalized by the likelihood fit to the observed dijet 
mass distribution with ${\cal B}(t \rightarrow H^+ b)$ 
fixed to 0.

The sources of systematic uncertainty in the extracted
${\cal B}(t \rightarrow H^+ b)$  
include uncertainties in the jet energy scale 
corrections~\cite{jes}, initial state and final-state 
radiation, modeling of the non-$t\bar{t}$ 
background, choice of event generators 
in simulation.
These systematic sources perturb the shape of 
the dijet mass and cause a shift in the result of 
the LH fit. The shift in the resulting
${\cal B}(t \rightarrow H^+ b)$ 
is estimated using ``pseudoexperiments'' of
the perturbed and unperturbed dijet mass distributions
for each systematic source; the pseudoexperiments 
are generated by the bin-to-bin Poisson fluctuations of 
the simulated dijet mass distributions.
The dominant systematic uncertainty
originates from the choice of event generators 
in the simulation, unless $m_{H^+}$ is close to 
$m_{W}$, in which case the jet energy scale 
uncertainty dominates. The other systematic uncertainties 
from data/Monte Carlo differences in $b$-tagging rates 
and top quark mass constraints in $t\bar{t}$ reconstruction 
are negligible compared to the uncertainties 
from the perturbed dijet mass shape. 

The individual systematic uncertainties are 
combined in quadrature. The total systematic 
uncertainty ($\Delta {\cal B}(t \rightarrow H^+ b)$) 
is represented by a nuisance parameter which adds to 
the branching ratio and has a Gaussian prior PDF
with width $\Delta {\cal B}(t \rightarrow H^+ b)$. 
We eliminate this nuisance
parameter by Bayesian marginalization~\cite{bayesian} and obtain 
a posterior PDF in ${\cal B}(t \rightarrow H^+ b)$ 
assuming a uniform prior PDF in 0 
$\leq {\cal B}(t \rightarrow H^+ b) \leq$ 1.
The expected upper limits on
${\cal B}(t \rightarrow H^+ b)$ with 95\% C.L.
are derived from a thousand
pseudoexperiments using the SM backgrounds 
events for each $m_{H^+}$. 

\begin{figure}[ht]
\includegraphics[width=3.375in]{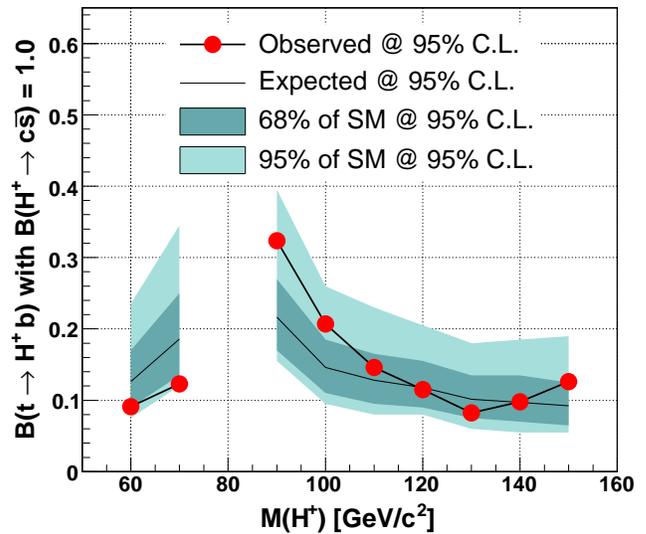}
\caption{\label{fig:cl95}The upper limits on 
${\cal B}(t \rightarrow H^+ b)$ at 95\% C.L for 
charged Higgs masses of 60 
to 150 GeV/$c^2$ except a region for 
$m_{H^+} \approx m_W$. The observed limits 
(points) in 2.2 ${\rm fb^{-1}}$ CDF II 
data are compared to the expected limits 
(solid line) with 68\% and 95\% uncertainty band.}
\end{figure}

The upper limits on ${\cal B}(t \rightarrow H^+ b)$ 
at 95\% C.L. show a good agreement between 
the observation and the SM expectation.
The upper limits in Figure~\ref{fig:cl95} includes
the systematic uncertainty in ${\cal B}(t \rightarrow H^+ b)$. 
Since the LH fit has very little sensitivity for 
$m_{H^+} \approx m_W$, the upper limits 
around 80 GeV/$c^2$ $H^+$ are omitted in the Figure.  
The exact values of the upper limits in the Figure~\ref{fig:cl95}
are listed in Table~\ref{tab:lim}. 

\begin{table}[htbp]
\begin{ruledtabular}
\begin{tabular}{c|ccccccccc}
$m_{H^+}$(GeV/$c^2$) & 60 & 70 & 90 & 100 & 110 & 120 & 130 & 140 & 150\\
\hline
Expected & 0.13 & 0.19 & 0.22 & 0.15 & 0.13 & 0.12 & 0.10 & 0.10 & 0.09 \\
Observed & 0.09 & 0.12 & 0.32 & 0.21 & 0.15 & 0.12 & 0.08 & 0.10 & 0.13 \\
\end{tabular}
\end{ruledtabular}
\caption{\label{tab:lim}Expected and Observed 
95\% C.L. upper limits on ${\cal B}(t \rightarrow H^+ b)$ 
for $H^+$ masses of 60 to 150 GeV/$c^2$. }
\end{table}

This analysis can set model-independent 
limits for anomalous scalar charged bosons production 
in top quark decays. Besides the assumption that a 
scalar boson decays only to $c\bar{s}$ with zero width, 
no model-specific parameter is used in this 
analysis. Therefore any generic charged boson
would make a secondary peak in the  dijet mass spectrum 
if it decays into a dijet final state like the 
$H^+ \rightarrow c\bar{s}$ in top quark decays.
Here, we extend the search below the $W$ boson 
mass~\cite{lep} down to 60 GeV/$c^2$ for any non-SM 
scalar charged boson produced in top quark decays,
$t \rightarrow X^+(\rightarrow u\bar{d}) b$. 
This process is simulated for the CDF II detector and 
is similar to $H^+ \rightarrow c\bar{s}$. 
In the simulation, we obtain a better
dijet mass resolution for $u\bar{d}$ decays than 
for the $c\bar{s}$ decays. 
The difference in the mass resolution comes from the 
smaller chance of false $b$-tagging from light quark final 
states of $X^+$ than the $c\bar{s}$ decays, thus
resulting in a smaller ambiguity of jet-parton assignments 
in the $t\bar{t}$ reconstruction. Consequently, 
the upper limits on 
${\cal B}(t \rightarrow X^+(\rightarrow u\bar{d}) b)$ 
are lower than the limits on 
${\cal B}(t \rightarrow H^+(\rightarrow c\bar{s}) b)$
regardless of the charged boson mass.

In summary, we have searched for a non-SM scalar
charged boson,
primarily the charged Higgs boson predicted in the 
MSSM, in top quark decays using lepton+jets $t\bar{t}$
candidates. This is the first attempt to search for  
$H^{+} \rightarrow c\bar{s}$ using fully reconstructed 
charged Higgs bosons. In the CDF II data sample of 
2.2 ${\rm fb^{-1}}$, we find no evidence of 
charged Higgs bosons in the dijet mass spectrum
of the top quark decays. Hence, upper limits on 
${\cal B}(t \rightarrow H^+ b)$ 
with 95\% C.L. are placed at 0.1 to 0.3 assuming
of ${\cal B}(H^+ \rightarrow c\bar{s})$ = 1.0 
for charged Higgs masses of 60 to 150 GeV/$c^2$. 
This analysis also yields conservative upper limits on 
any non-SM scalar charged boson $X^+$ production from 
top quarks. 
Based on simulation, we find that the upper limits on the 
branching ratio ${\cal B}(X^+ \rightarrow u\bar{d})$ are always
better than the upper limits on ${\cal B}(H^+ \rightarrow c\bar{s})$.

We thank the Fermilab staff and the technical staffs of the participating institutions for their vital contributions. This work was supported by the U.S. Department of Energy and National Science Foundation; the Italian Istituto Nazionale di Fisica Nucleare; the Ministry of Education, Culture, Sports, Science and Technology of Japan; the Natural Sciences and Engineering Research Council of Canada; the National Science Council of the Republic of China; the Swiss National Science Foundation; the A.P. Sloan Foundation; the Bundesministerium f\"ur Bildung und Forschung, Germany; the Korean Science and Engineering Foundation and the Korean Research Foundation; the Science and Technology Facilities Council and the Royal Society, UK; the Institut National de Physique Nucleaire et Physique des Particules/CNRS; the Russian Foundation for Basic Research; the Ministerio de Ciencia e Innovaci\'{o}n, and Programa Consolider-Ingenio 2010, Spain; the Slovak R\&D Agency; and the Academy of Finland.


\begin{thebibliography}{1000}
\bibitem{tstar}
The $t^*$ is a virtual top quark with an off-shell mass.
\bibitem{w}
The positive and negative charged weak boson 
($W^{\pm}$) is presented as $W$ in this Letter.
\bibitem{lepton}
In this Letter, the word lepton ($l$) stands for an 
electron or a muon. The tau lepton is not referred to 
as a lepton, but is specified as $\tau$.
\bibitem{eta}
The detector uses cylindrical coordinates where
$\theta$ is the polar angle with respect to the proton beam
and $\phi$ is the azimuthal angle. The direction of a 
particle in the detector is expressed with the 
pseudorapidity, $\eta = -\ln \tan(\theta/2)$.
\bibitem{met}
Transverse energy is $E_T = E\sin\theta$, where $E$ is the 
measured energy. Transverse momentum, $p_T$, of charged
particles, is calculated similarly. From energy-momentum
conservation, $\sum_i E^i_T \hat{n_i} = 0$ for radial unit
vector $\hat{n_i}$ in the azimuthal plane. Missing transverse 
energy is defined as $- \sum_i E^i_T \hat{n_i}$, representing 
undetected particles such as neutrinos. The missing $E_T$ 
is further corrected for the energy and momentum of 
identified muons.
\end{thebibliography}
\end{document}